\documentclass[12pt]{article}
\usepackage{amsmath,amsfonts,amssymb,setspace}
\usepackage[utf8]{inputenc}
\usepackage[nosort]{cite}
\usepackage[usenames,table]{xcolor}
\usepackage{dsfont}
\usepackage{mathrsfs}
\usepackage{graphicx}	
\usepackage{float}
\usepackage{longtable}	
\usepackage{lipsum}
\usepackage{verbatim}
\usepackage{leftidx}	
\usepackage{bm}			
\usepackage{marginnote}	

\numberwithin{equation}{section} 

\usepackage[top=2.8cm,
			bottom=2.8cm,
			left=2.0cm,
			right=2.0cm]{geometry}
\usepackage[colorlinks=true,
			linkcolor=black,
			citecolor=black,
			urlcolor=blue,
			filecolor=black]{hyperref}

\allowdisplaybreaks

\begin{document}

\vspace{30pt}

\begin{center}


{\Large\sc Fractional Particle with Fractional First Derivatives 
\\[12pt]}

\vspace{-5pt}
\par\noindent\rule{495pt}{0.5pt}


\vskip 1cm

{\sc A. V. Crisan$^{a}$, C. M. Porto$^{b}$, C. F. L. Godinho$^{b}$ and I. V. Vancea$^{b}$ }


\vspace{10pt}
{\it 
$^{a}$Department of Mechanical Systems Engineering,\\
Technical University of Cluj-Napoca,\\
103 – 105 Muncii Bld., C.P. 400641 Cluj-Napoca, Romania
\\
$^{b}$Group of Theoretical Physics and Mathematical Physics,\\
Department of Physics, Federal Rural University of Rio de Janeiro,\\
Cx. Postal 23851, BR 465 Km 7, 23890-000 Serop\'{e}dica - RJ,
Brazil}

\vspace{4pt}


{\tt\small
\href{mailto:adina.crisan@mep.utcluj.ro}{adina.crisan@mep.utcluj.ro};
\href{mailto:claudio@ufrrj.br}{claudio@ufrrj.br};
\href{mailto:crgodinho@gmail.com}{crgodinho@gmail.com}; 
\href{mailto:ionvancea@ufrrj.br}{ionvancea@ufrrj.br}
}


\vspace{30pt} {\sc\large Abstract}

\end{center}

In this paper, we introduce a new classical fractional particle model incorporating fractional first derivatives. This model represents a natural extension of the standard classical particle with kinetic energy being quadratic in fractional first derivatives and fractional linear momenta, similarly to classical mechanics. We derive the corresponding equations of motion and explore the symmetries of the model. Also, we present the formulation in terms of fractional potentials. Two important examples are analytically solved: the free particle and the particle subjected to generalized forces characterized by fractional first derivatives.

\noindent



\newpage


\section{Introduction}

By combining non-local dynamics generated by the one-dimensional fractional Laplacian pseudo-differential operator with the classical principle of least action, classical fractional particle models that generalize standard classical particles were constructed in \cite{Vancea:2023tpb,Vancea:2023cc}. These models are characterized by a non-quadratic kinetic term of the form $\sim x_{a}(t) (- \Delta_t)^{\frac{\alpha}{2}} x^{a}(t)$, where $x^{a}(t)$ are the particle's instantaneous coordinates in the $n$-dimensional Euclidean space and $(- \Delta_t)^{\frac{\alpha}{2}}$ is the one-dimensional fractional Laplacian in time variable. The main motivation for studying classical fractional particles with a fractional Laplacian in the kinetic term is the low-dimensional realization of the Caffarelli-Silvestre extension problem \cite{Caffarelli:2007}. In this framework, the equivalence between the one-dimensional and two-dimensional classical actions is interpreted as a mapping between the fractional particle with a fractional time-like Laplacian on the boundary of the two-dimensional sigma model defined on the half-plane, and the two-dimensional field theory defining the sigma model \cite{Vancea:2023tpb}.

These fractional particle models have various applications. The fractional Laplacian is important for constructing more general models of diffusion and random walk processes and in studying systems with long-range interactions and correlations. A well know application is the use of classical fractional particle models in studying fractional Brownian motion, which exhibits long-range correlations in statistical quantities \cite{Bock:2020}. Other potential applications include modeling systems with viscoelastic properties, such as certain polymers and soft tissues, which are difficult to describe using standard classical mechanics. In these systems, particle dynamics are influenced by memory effects and non-local interactions, which can be formulated within the fractional Laplacian formalism \cite{Failla:2020}. This formalism also applies to heat conduction in materials with anomalous thermal behavior, such as fractal structures or materials with long-range correlations \cite{Zecova:2015}, and turbulent flows \cite{Suzuki:2023}, where classical fractional particles can provide detailed system descriptions. Also, the classical fractional particle could provide a better understanding of fractional diffusion in porous media, providing tools for studying of particle diffusion through porous structures with long-range interactions and fractional derivatives \cite{Vasquez:2012}.

It should be mentioned that fractionality can also be introduced at the level of the equations of motion rather than the action. Classical relativistic fractional particles, whose equations of motion are written in terms of Caputo fractional derivative, were previously developed as generalizations of the relativistic particle with linear dissipation \cite{Tarasov:2010,Gonzalez:2007} and the classical Dirac particle with memory \cite{Tarasov:2020}. The generalization of classical oscillators using Caputo fractional time derivatives is discussed in \cite{Nasrolahpour:2012,Gomez-Aguilar:2012}.

In the fractional particle model with the fractional Laplacian, the kinetic term is not quadratic in the derivatives of $x^{a}(t)$, complicating the understanding of linear momentum. As noted in \cite{Vancea:2023tpb}, the kinetic energy arises from the interaction between the standard local momentum and the canonical non-local momentum defined by the fractional first derivative 
$ \left(-\Delta^{(1)}_{t}\right)^{\frac{\alpha}{2}}$. This leads to the question of whether there exists a formulation of the fractional particle where the kinetic energy is quadratic in the fractional linear momentum, similar to standard classical mechanics.

In this paper, we address this question by constructing and analyzing a fractional particle model where the kinetic term is defined in terms of the fractional first derivative. This model does not directly connect with field theories via the extension problem, which requires the use of the fractional Laplacian, and it is distinct from the previously studied fractional Laplacian particle model from \cite{Vancea:2023tpb}.

The fractional particle model with the kinetic term defined as $(- \Delta^{(1)}_t)^{\frac{\alpha}{2}} x_{a}(t) (-\Delta^{(1)}_t)^{\frac{\alpha}{2}} x^{a}(t)$ has specific properties and applications. One potential application is in the study of anomalous diffusion processes where the underlying mechanisms are best described by fractional derivatives, providing a more accurate representation of memory and hereditary properties in complex systems \cite{Metzler:2000}. 
Another interesting application is in signal processing, where fractional derivatives could contribute to the analysis of signals with fractal-like properties, or in systems with non-local interactions \cite{Debnath:2003}. 
Also, this model could be applied to complex media, such as heterogeneous materials or porous structures, where traditional kinetic term fails to fully capture the particle dynamics \cite{Sokolov:2012}. Other potential applications lie in financial mathematics, where asset price movements exhibit long-range dependence \cite{{RamaCont:2001}} and in biological systems, where the model can describe the movement of cells or organisms in environments where traditional models are inadequate, such as in the context of tissue engineering or the spread of diseases with non-local interaction patterns \cite{Mendez:2010}. 

Both the fractional Laplacian and the fractional first derivative are non-local operators that integrate over the entire domain, employing function values from a broad range of points and being applicable to similar phenomena. Therefore, it is natural to ask how fractional particle models build with these two operators provide distinct understandings of the system under study. A principal distinction between the models arises from the relationship between kinetic energy and canonical momentum, leading to important differences in both formal aspects and physical interpretations. Another distinction is evident in how each operator addresses the non-locality. While the fractional particle with fractional Laplacian captures long-term dependencies by integrating differences over the entire temporal axis, the fractional particle with fractional first derivative models temporal memory and hereditary properties by considering the function values symmetrically around $t$, emphasizing local values rather than differences. More detailed distinctions are phenomenon-dependent and can only be observed through the study of concrete applications.

This paper is organized as follows. Section 1 introduces the concept of a fractional particle characterized by a fractional first derivative. We establish the kinetic term by substituting standard derivatives with fractional operators and derive the equation of motion, also addressing the symmetries of the action. Section 2 formulates the fractional particle with fractional first derivative using potentials. Section 3 presents two examples: the free fractional particle and one subjected to a generalized force with a fractional first derivative. In the final section, we discuss our findings. Appendix A provides the definition and essential properties of the fractional first derivative used throughout this paper. Appendix B contains the proof of an integral result referenced in the main text.

\section{Fractional Particle with Fractional First Derivative} 
\label{sec:FPM}

The construction of models with fractional operators follows the general principle of replacing the differential operators from standard calculus with the  corresponding pseudo-differential operators from fractional calculus. Since the orders of operators in the two formalisms differs from each other, additional criteria must be chosen to construct the kinetic term, and potentially in the interaction terms as well. When applied to the fractional particle, this principle can be implemented by requiring that the kinetic term be quadratic in fractional operators, similar to the classical case. Other possibilities may arise, justified by physical or mathematical conditions, as discussed, for example, in \cite{Vancea:2023tpb,Vancea:2023cc}.

Consider a classical particle of mass $m$ moving in the Euclidean space under the influence of the potential $V[x]$, where $\{ x^{a} \}_{a = \overline{1,n}}$ are the components of vector $x \in \mathbb{R}^n$ in Cartesian coordinates. We write the action functional as
\begin{equation}
S[x] =\int dt \left[ \frac{m}{2} \frac{d x_{a} (t)}{d t} \frac{d x^{a} (t)}{d t} - V[x] \right]
\, ,
\label{FPFD-act-stand}
\end{equation}
where we use covariant notation and Einstein summation for convenience.

We propose a classical fractional particle model that generalizes the action (\ref{FPFD-act-stand}) in the following way. At any instant $t \in \mathbb{R}$, the fractional particle is localized at $x^{a} (t)$ which depends on all positions $x^{a}(t^{\prime})$ with $t^{\prime} \in I \subseteq \mathbb{R}$. For example, this is the case of a system with memory and $n$-degrees of freedom in local or global interaction. Various pseudo-operators of fractional calculus can be used to express the kinetic energy of the fractional particle \cite{Tarasov:2010book,Anastassiou:2022}. However, as we mentioned in introduction, here we are interested in constructing a model with kinetic term quadratic in the fractional first derivative $\left(-\Delta_t^{(1)}\right)^{\frac{\alpha}{2}}$ which should be used instead of $d/d t$ according to the general principle. The fractional particle action is given by 
\begin{equation}
S^{(\alpha)}[x] = \int_{-\infty}^{+\infty} dt \left[ \frac{m_{\alpha}}{2} 
\left( - \Delta_t^{(1)}\right)^{\frac{\alpha}{2}} x_{a} (t)
\left( - \Delta_t^{(1)}\right)^{\frac{\alpha}{2}} x^{a} (t)
 - V[x] \right]
\, ,
\label{FPFD-act}
\end{equation}
where the potential term is local. The fractional first derivative from the equation (\ref{FPFD-act}) is defined by
\begin{equation}
\left(-\Delta_t^{(1)}\right)^{\frac{\alpha}{2}} f(t)
= c_\alpha^{(1)} \text{P.V.} \int_{-\infty}^{+\infty} \frac{\tau f(t+\tau)}{|\tau|^{1+\alpha}} 
d \tau
\, ,
\label{FPFD-der-1}
\end{equation}
where $\alpha \in [0, 2)$ and the constant is expressed in terms of Euler Gamma functions
\begin{equation}
c_\alpha^{(1)}=\frac{2^{\alpha-1}}{\sqrt{\pi}} \frac{\Gamma\left(\frac{1+\alpha}{2}\right)}{\Gamma\left(\frac{2-\alpha}{2}\right)}
\, .
\label{FPFD-der-2}
\end{equation}
The principal value in the definition (\ref{FPFD-der-1}) is necessary to ensure a well defined expression at $\tau = 0$. In general, the functions $f(t)$ should be sufficiently smooth and exhibit appropriate decay behavior as $\lvert t \rvert$ approaches infinity. Examples of such spaces are Lesbegue spaces $\mathcal{L}^p$, Sobolev space $\mathcal{W}^{k,p}$, H\"{older} spaces $\mathcal{C}^{0,\alpha}$, Bessel potential spaces $\mathcal{H}^s$, etc. The behavior of $f(t)$ near zero is critical: if $f(t)$ has a singularity at $t = 0$, the fractional first derivative may not exist. Here, we denote the space of allowed functions by $\mathcal{X}$ and we refer to Appendix A for more details. 
The parameter $m_{\alpha}$ in $S^{(\alpha)}[x]$ has dimension $E^{5-2\alpha}$ in natural units $\hbar=c=1$. Thus, in the upper limit of fractionality parameter $\alpha$ we have \cite{Pozrikidis:2016}
\begin{equation}
\lim_{\alpha \to 2} m_{\alpha}=m
\, ,
\qquad
\lim_{\alpha \to 2} \left(-\Delta_t^{(1)}\right)^{\frac{\alpha}{2}} x^{a}(t) = \frac{d x^{a} (t)}{d t}
\, , 
\label{FPFD-lim}
\end{equation}
which defines the standard classical particle limit of the model
\begin{equation}
\lim_{\alpha \to 2} S^{(\alpha)} [x] = S[x]
\, .
\label{FPFD-lim-S}
\end{equation}
Observe that, as in the case of classical fractional particles with fractional Laplacian, the time parameter in the action $S^{(\alpha)}[x]$ belongs to 
$\mathbb{R}$. From a mathematical point of view, this allows the exploitation of the properties of the fractional first derivative to derive the equations of motion and to discuss the symmetries of the action. From a physical perspective, this can be interpreted as the contribution of trajectories defined for all values of time to the action, a consequence of system's non-locality in time.  

The model can be extended to include non-local interaction terms, for example,
\begin{equation}
V[x](t) = \int d t^{\prime} K_a(t,t^{\prime}) x^{a} (t^{\prime})  
\, .
\label{FPFD-int-nonloc}
\end{equation}
While these extensions are very interesting and natural for certain applications, we will not discuss them here. Our focus is on understanding the effect of the fractional first derivative on particle dynamics, which is contained in the kinetic energy term.

\subsection{Equations of Motion}\label{sub:eq-motion}

To derive the equations of motion, we apply the variational principle to the action $S^{(\alpha)}[x]$. We consider small variations in the coordinates \cite{Vancea:2023tpb,Vancea:2023cc}
\begin{equation}
x_{a} \left( t \right) \mapsto x^{\prime}_{a} \left( t \right) = x_{a} \left( t \right) + \delta x_{a} \left( t \right)
\, ,
\label{FPFD-eq-mot-var}
\end{equation}
with $\delta x^{a} (t) \to 0$ as $\lvert t \rvert \to \infty$. The variation of the action is then given by
\begin{align}
\delta S^{(\alpha)} [x] 
& = S^{(\alpha)} [x + \delta x] - S^{(\alpha)} [x]
\nonumber
\\
& \simeq \frac{m_{\alpha}}{2} \int_{-\infty}^{+\infty} dt 
\left[ 
	\left( - \Delta_t^{(1)}\right)^{\frac{\alpha}{2}} \delta x_{a} (t)
	\left( - \Delta_t^{(1)}\right)^{\frac{\alpha}{2}} x^{a} (t)
\right.
\nonumber
\\
& \quad +
\left.
	\left( - \Delta_t^{(1)}\right)^{\frac{\alpha}{2}} x_{a} (t)
	\left( - \Delta_t^{(1)}\right)^{\frac{\alpha}{2}} \delta x^{a} (t)
	- \frac{\partial V[x]}{\partial x^{a}(t)}  \delta x^{a}(t)
\right]
\, .
\label{FPFD-eq-mot-act} 
\end{align}
Using the identity proved in \cite{Vancea:2023cc},
\begin{equation}
\int_{-\infty }^{+\infty } dt \, f(t) \left( -\Delta_t^{(1)} \right)^{\frac{\alpha}{2}} g(t)
=
- \int_{-\infty }^{+\infty } dt \, g(t) \left( -\Delta_t^{(1)} \right)^{\frac{\alpha}{2}} f(t)
\label{FPFD-eq-mot-id}
\end{equation}
we obtain the following equations of motion
\begin{equation}
m_{\alpha} \left( -\Delta_t^{(1)} \right)^{\frac{\alpha}{2}} \left( -\Delta_t^{(1)} \right)^{\frac{\alpha}{2}} x_{a} (t) + \frac{\partial V[x]}{\partial x^{a}(t)} = 0
\, ,
\label{FPFD-eq-mot-gen}
\end{equation} 
for all $a = 1, \ldots , n$.

Some comments are in order here. Firstly, we observe that the potential term remains the same as in classical mechanics due to its local nature. However, if non-local interactions are considered, the potential term must be modified accordingly. Secondly, the term resulting from the equations of motion involves two consecutive actions of the fractional first derivative on the coordinate function. In the classical limit $\alpha \to 2$, this term reduces to the familiar second derivative $d^2/dt^2$, providing a consistency check for equation (\ref{FPFD-eq-mot-gen}). 

Due to the non-local nature of the operator $\left( -\Delta_t^{(1)} \right)^{\frac{\alpha}{2}}$, the equations of motion become double singular integral equations, which are generally more complex to solve than classical differential equations. To appreciate their mathematical structure, we rewrite equation (\ref{FPFD-eq-mot-gen}) using the definition (\ref{FPFD-der-1})
\begin{equation}
m_\alpha \left[c_\alpha^{(1)}\right]^2 
\int_{-\infty}^{+\infty} \frac{d \tau \, \tau}{|\tau|^{1+\alpha}}
\left[\int_{-\infty}^{+\infty} \frac{d \zeta \, \zeta}{|\zeta|^{1+\alpha}} x_a (t+\tau+\zeta)\right] + \frac{\partial V[x]}{\partial x^{a}(t)} = 0
\, .
\label{FPFD-eq-mot-int}
\end{equation}

As in classical mechanics, the equations of motion can be generalized to include external forces
\begin{equation}
m_{\alpha}
\left( -\Delta_t^{(1)} \right)^{\frac{\alpha}{2}}
\left( -\Delta_t^{(1)} \right)^{\frac{\alpha }{2}}
x_a (t) + \frac{\partial V[x]}{\partial x^{a}(t)} = f_a (x,t)
\, ,
\label{FPFD-eq-mot-gen-1}
\end{equation} 
where $f_a (x,t)$ are the components of the generalized force $f(x,t)$. These generalized forces are expected to play a significant role in the description of systems involving fractional particles, as they can more accurately model the phenomena to which fractional calculus is applied.

\subsection{Symmetries}\label{sub:symm}

We now turn to a general analysis of symmetries of the action $S^{(\alpha)}[x]$. We start off by observing that the group $SO(n)$ of rotations in 
$\mathbb{R}^n$ acts on the vectors $x$ as
\begin{equation}
x^{a} \to x^{\prime \, a} = {R^{a}}_{b} \, x^{b}
\, ,
\label{FPFD-sym-rot-1}
\end{equation} 
where ${R^{a}}_{b}$ belongs to the defining representation of $SO(n)$. Then it is straightforward to see that the kinetic term is invariant under the action of rotations since
\begin{equation}
\left( - \Delta_t^{(1)}\right)^{\frac{\alpha}{2}} x^{\prime}_{a} (t)
\left( - \Delta_t^{(1)}\right)^{\frac{\alpha}{2}} x^{\prime \, a} (t)
=
\left( - \Delta_t^{(1)}\right)^{\frac{\alpha}{2}} x_{a} (t)
\left( - \Delta_t^{(1)}\right)^{\frac{\alpha}{2}} x^{a} (t)
\, .
\label{FPFD-sym-rot-2}
\end{equation}
The invariance of the potential term is ensured by the condition $V[x]=V[x^2]$ which is the same as in classical mechanics. This defines systems that are rotational invariant. 

Consider next the translations in $\mathbb{R}^n$ given by
\begin{equation}
x^a (t) \to x^{\prime \, a} (t) = x^{a} (t) + u^{a}
\, ,
\label{FPFD-sym-tran-1}
\end{equation}
where $u^{a}$ are the components of a constant vector $u \in \mathbb{R}^n$. Under the transformations (\ref{FPFD-sym-tran-1}), the kinetic term transforms as
\begin{equation}
K_{\alpha}[x] \to K_{\alpha}[x'] = K_{\alpha}[x] + 2 I_{\alpha}
u_{a} \left( - \Delta_t^{(1)}\right)^{\frac{\alpha}{2}} x^{a} (t)
+ u^2 I^{2}_{\alpha}
\, ,
\label{FPFD-sym-tran-2}
\end{equation}
where
\begin{equation}
I_{\alpha} = 
\text{P.V.} \int_{-\infty}^{+\infty} \frac{d \tau \, \tau}{|\tau|^{1+\alpha}}
\, .
\label{FPFD-sym-tran-3}
\end{equation} 
The principal value integral $I_{\alpha}$ vanishes for $\alpha \in (1,2)$ as proved in Appendix B. It follows that the kinetic term $K_{\alpha}[x]$ is translation invariant for values of the fractionality parameter greater than one. However, for $\alpha \leq 1$, the translation invariance is broken. At $\alpha = 1$, the kinetic energy becomes divergent. Therefore, the action $S^{(\alpha)}[x]$ is translationally invariant if $\alpha \in (1,2)$ and the potential $V[x]$ is translationally invariant as well, as is the case with the linear oscillator or Coulomb potential.

Another symmetry encountered in classical mechanics is the invariance under time reparametrization. In general, this is not a symmetry for the fractional particle since, unlike the first derivative $d/dt$, the fractional first derivative is not invariant under time reparametrization. This can be easily seen by considering the transformation
\begin{equation}
t \to t^{\prime} = t + \theta
\, ,
\label{FPFD-sym-time-1}
\end{equation}
where $\theta$ is a real constant. Under the transformation (\ref{FPFD-sym-time-1}), the fractional first derivative transforms as
\begin{equation}
\left( - \Delta_t^{(1)}\right)^{\frac{\alpha}{2}} x_{a} (t)
\to
\left( - \Delta_{t^{\prime}}^{(1)}\right)^{\frac{\alpha}{2}} x_{a} (t^{\prime})  
=\left(-\Delta_t^{(1)}\right)^{\frac{\alpha}{2}} x_{a} (t)+\sum_{s=1}^{\infty} \frac{\theta^s}{s!}\left(-\Delta_t^{(1)}\right)^{\frac{\alpha}{2}} x_{a}^{(s)}(t)
\, ,
\label{FPFD-sym-time-2}
\end{equation}
where $x_{a}^{(s)}(t)$ is the standard $s$-derivative of $x_{a}(t)$. This is intuitive since the system's instantaneous position depends on all past and future states. Nevertheless, if one restricts the space of $x(t)$ to time-invariant functions, the action remains time-invariant for stationary potentials. 

We conclude this subsection with some general remarks on the kinetic energy $K_{\alpha}[x]$ of the fractional particle. By construction, $K_{\alpha}[x]$ is quadratic in the fractional first derivatives of the coordinates according to (\ref{FPFD-act}). This suggests interpreting these derivatives as the components of fractional linear momentum of the particle
\begin{equation}
p^{(\alpha )}_{a} (t) = m_{\alpha} \left( - \Delta_t^{(1)}\right)^{\frac{\alpha}{2}} x_{a} (t)
\, .
\label{FPFD-lin-mom-def}
\end{equation} 
In the classical limit, this becomes
\begin{equation}
\lim_{\alpha \to 2} p^{(\alpha )}_{a} (t) = p_{a} (t) = m \frac{d x_{a} (t)}{d t}
\, .
\label{FPFD-lin-mom-1}
\end{equation}
Thus, the kinetic energy of a fractional classical particle in the fractional first derivative model takes the form
\begin{equation}
K_{\alpha} [x] = \frac{p^{(\alpha )}_{a} p^{(\alpha )\, a}}{2 m_{\alpha}}
\, ,
\qquad
\lim_{\alpha \to 2 } K_{\alpha} [x] = K[x] = \frac{p_a p^a}{2 m}
\, .
\label{FPFD-lin-mom-2}
\end{equation}
Both the kinetic energy and momentum of the fractional particle are non-local quantities in time for all permissible values of the fractionality parameter $\alpha$, depending on the particle's states at all values of $t$.

\subsection{Formulation in Terms of Laplacian Potentials}\label{sub:pot}

The fractional particle model defined in terms of the fractional Laplacian can be reformulated in terms of Laplacian potentials, where the kinetic energy is expressed through the standard second derivative \cite{Vancea:2023tpb, Vancea:2023cc}. The fractional particle model with a fractional first derivative admits a similar reformulation. To this end, we introduce a Laplacian potential $X^{a}(t)$ for each coordinate $x^{a}(t)$ by the relation
\begin{equation}
X^{a}(t)=c^{(\alpha)}_{0}
\text{P.V.}\int_{-\infty}^{\infty} \frac{x^{a}(t+\tau)}{\lvert \tau \rvert^{1-\alpha}} d\tau 
\, ,
\qquad
c^{(\alpha)}_{0} = \frac{2^{\alpha - 1}}{\sqrt{\pi}}\frac{\Gamma \left(\frac{\alpha -1}{2}\right)}{\Gamma\left(\frac{2-\alpha}{2} \right)}
\, ,
\label{FPFD-lap-1}
\end{equation}
where $c^{(\alpha)}_{0}$ is a constant determined by the Euler Gamma functions. Using this potential, we can rewrite the fractional particle action as
\begin{equation}
S^{(\alpha)}[X] = \int_{-\infty}^{\infty} dt
\left[ M_{\alpha} \frac{d X_{a}(t)}{d t} \frac{d X^{a}(t)}{d t}
- V[X]
\right]
\, .
\label{FPFD-lap-2}
\end{equation}

This reformulation shows the equivalence between the kinetic energy term and a kinetic term expressed in terms of Laplacian potentials. The mass parameter $M_{\alpha}$ in this context is related to the original mass parameter $m_{\alpha}$ and the constant $c^{(\alpha)}_{0}$ through the relation
\begin{equation}
M_{\alpha} = m_{\alpha} \left( c^{(\alpha)}_{0} \right)^2
\, .
\label{FPFD-lap-3}
\end{equation}

The potential $V[X]$ keeps the same functional form but now depends on the Laplacian potentials $X^{a}(t)$. This transformation can simplify the analysis of the system, as it translates the non-local fractional derivative terms into more familiar local terms involving standard derivatives. Furthermore, the reformulation in terms of Laplacian potentials can facilitate the application of standard techniques from classical mechanics and quantum mechanics to the study of fractional particle dynamics.

Formally, the equations of motion for the Laplacian potentials are analogous to those for standard coordinates. However, the action (\ref{FPFD-lap-2}) differs from the standard particle action in certain crucial aspects. Firstly, the mass parameter $ M_{\alpha} $ has dimensions $ E^{1+2\alpha} $, highlighting its fractional nature. Secondly, the potential term $ V[X] $ now depends on the Laplacian potentials $ X $ rather than the coordinates $ x $, leading to a new class of potentials that describe non-local interactions in time. Thirdly, while the kinetic energy resembles that of a standard particle and the formal momentum components $ P^{\alpha}_{a} (t) = M_{\alpha} \frac{d X_{a}(t)}{dt} $ are defined, these quantities remain non-local in time and lack a straightforward geometrical interpretation unlike in classical mechanics. Nevertheless, replacing the fractional first derivative with the standard derivative allows us to reformulate the equations of motion from integral equations to differential equations for the potentials.

To derive the equations of motion, we apply the variational principle to the action $ S^{(\alpha)}[X] $. The variation of the functions $ X^{a}(t) $ must be consistent with their definition as Laplacian potentials
\begin{equation}
\delta X^{a}(t)=c^{(\alpha)}_{0}
\text{P.V.}\int_{-\infty}^{\infty} \frac{\delta x^{a}(t+\tau)}{\lvert \tau \rvert^{1-\alpha}} d\tau 
\, .
\label{FPFD-lap-3-1}
\end{equation}
This definition implies that the interaction term $ V[X] $ depends on potentials rather than coordinates. Another consequence is that, to apply the variational principle rigorously, we must work within the set of functions $ \mathcal{D} $ defined by
\begin{equation}
\mathcal{D} = \left\{ f \in \mathcal{X} \, : \, \forall \epsilon > 0, 
\exists \,\, \delta > 0 \,\, \text{such that} \,\, \lvert \delta X^{a} (t)\rvert < \delta \,\, \text{if} \,\, \lvert \delta x^{a} (t) \rvert < \epsilon ,  \,\,
\forall t \in \mathbb{R}
\right\}
\, ,
\label{FPFD-lap-4}
\end{equation}
where $ \mathcal{X} $ denotes the function space used in the fractional particle model. Consequently, the equations of motion take on the familiar form
\begin{equation}
M_{\alpha} \frac{d^2 X_{a}(t)}{d t^2}+\frac{\partial V[X]}{\partial X_{a}} = f_{a}[X,t]
\, ,
\label{FPFD-lap-5}
\end{equation}
where $ f_{a}[X,t] $ represents the generalized forces that also depend on the potentials. For problems involving Laplacian potentials, solutions to (\ref{FPFD-lap-5}), subject to appropriate boundary conditions, provide  solutions of the problem as usual. However, if the coordinates $ x^{a}(t) $ are also required, obtaining the final solution involves inverting relation (\ref{FPFD-lap-1}), generally necessitating computer-assisted numerical methods due to its non-trivial nature.

\section{Examples of Fractional First Derivatives Particles} 
\label{sec:examples}

In this section, we discuss some particular cases of the model presented in the previous section in which analytic solutions of the equations of motion can be found. In more general situations, due to the nature of these equations as double integral non-local equations, computational and numerical methods are essential for finding solutions in more complex cases.

\subsection{Free Particle}\label{sub:fp}

The simplest case to consider is that of a particle moving freely, that is with $V[x]=0$ and $f(t,x)=0$. Under these assumptions, the equations of motion (\ref{FPFD-eq-mot-gen-1}) simplify to
\begin{equation}
m_{\alpha}
{\left( -\Delta _{t}^{(1)} \right)}^{\frac{\alpha}{2}}
{\left( -\Delta _{t}^{\left( 1 \right)} \right)}^{\frac{\alpha }{2}}
{{x}_{a}} (t) = 0
\, .
\label{EX-fp-1}
\end{equation}
The simplest solution to the equation (\ref{EX-fp-1}) is a particle with zero fractional momentum $p^{(\alpha)}_{a} (t) = 0$ which, according to (\ref{FPFD-lin-mom-2}), also implies zero kinetic energy. Consequently, the coordinate time distributions $x^{a}(t)=x^{a}_{0}$ are constant. To determine the constraints on free particle solutions, we substitute the constant solution into the zero fractional momentum condition. Using equation (\ref{FPFD-der-1}), we obtain
\begin{equation}
\left(-\Delta_t^{(1)}\right)^{\frac{\alpha}{2}} x^{a}_{0}
= x^{a}_{0} c_\alpha^{(1)}  \text{P.V.} \int_{-\infty}^{+\infty} \frac{\tau d \tau}{|\tau|^{1+\alpha}}
d \tau  =  x^{a}_{0} c_\alpha^{(1)} I_{\alpha} = 0
\, .
\label{EX-fp-2}
\end{equation}
From equation (\ref{app:B-8}), we conclude that the constant coordinate time distributions are solutions to the equations of motion for $\alpha \in (1,2)$. For $\alpha \in (0,1]$, the equations of motion do not permit constant solutions because the integral $I_{\alpha}$ diverges. The trivial case $x^{a}_{0} = 0$ is a solution for all values of the fractionality parameter in the range $[0,2)$.

As we have seen in the previous section, in the limit $\alpha \to 2$, the fractional particle model converges to the standard particle model. This correspondence between the two models raises the question of the inertia law, 
specifically whether linear solutions to the equations of motion exist.
To demonstrate that the linear coordinates $x^{a}(t)=v^{a}t + x^{a}_{0}$, where $v^{a}$ are the components of the constant velocity vector, satisfy the equations of motion, we use the equation (\ref{app:A-5}) and obtain
\begin{align}
& m_{\alpha}
{\left( -\Delta _{t}^{(1)} \right)}^{\frac{\alpha}{2}}
{\left( -\Delta _{t}^{\left( 1 \right)} \right)}^{\frac{\alpha }{2}}
(v^{a}t + x^{a}_{0}) 
\nonumber
\\
& = 
\left[c^{(1)}_{\alpha}\right]^2 
\int_0^{+\infty} \frac{d \tau}{\tau^\alpha} 
	\left[ 
		\int_0^{+\infty} \frac{d \zeta}{\zeta^\alpha}
			\left[
				v^{a}(t+\tau+\zeta) - v^{a}(t+\tau-\zeta)
				- v^{a}(t-\tau+\zeta) + v^{a}(t-\tau-\zeta)
			\right]
	\right]
=0
\, .
\label{EX-fp-3}
\end{align}
The above result shows that the linear distributions are solutions of the equations of motion for all values of the fractionality parameter. Also, since we have used the full equations of motion in equation (\ref{EX-fp-3}), the 
fractional momentum is not-zero. It is easy to verify that for all $\alpha \in [0,2)$, the momentum diverges as $|t| \to \infty$ as does the kinetic energy. Introducing an energy cutoff $\Lambda$, the linear momentum and kinetic energy diverge as
\begin{equation}
p^{\alpha}_{a} = \frac{2 m_{\alpha} c^{(1)}_{\alpha} v_{a}}{2-\alpha} \Lambda^{\alpha - 2}
\, ,
\qquad
K_{\alpha} = \frac{2 m_{\alpha} [c^{(1)}_{\alpha}]^2 v^2}{(2-\alpha)^2} \Lambda^{2\alpha - 4}
\, .
\label{EX-fp-4}
\end{equation}
Equation (\ref{EX-fp-4}) suggests the following interpretation: to mantain a coordinate distribution linear in time, the fractional particle must increase its kinetic energy over time. Thus, even in the absence of external forces, the model points to an existing background interaction non-local in time modelled by the fractional first derivative.

\subsection{Particles under Generalized Forces}\label{sub:gf}

The effects modeled by kinetic terms with fractional first-order derivatives, some of which were mentioned in the introduction, could be generated by generalized forces rather than the usual potentials, which generally do not require fractional derivatives.

Consider a fractional particle under the influence of a non-local generalized force. According to the equations of motion (\ref{FPFD-eq-mot-gen-1}), the fractional momentum changes non-locally as
\begin{equation}
{\left( -\Delta _{t}^{(1)} \right)}^{\frac{\alpha }{2}}
{p}^{\alpha}_{a} (t) = f_{a} (x,t)
\, .
\label{EX-gf-1}
\end{equation} 
We are interested in generalized forces whose non-local effect is generated by the fractional first derivative from suitable local force currents
\begin{equation}
f_{a} (x,t) = {\left( -\Delta _{t}^{(1)} \right)}^{\frac{\alpha }{2}} F_{a} (x,t)
\, .
\label{EX-gf-2}
\end{equation} 
Substituting the right-hand side of equation (\ref{EX-gf-2}) into equation (\ref{EX-gf-1}) and applying the Fourier transform, we get
\begin{equation}
m_\alpha \mathcal{F} \left\{\left(-\Delta_t^{(1)}\right)^{\frac{\alpha}{2}} x_a (t) ; \omega\right\}
=
\mathcal{F}\left\{F_a(t, x) ; \omega\right\}
\, .
\label{EX-gf-3}
\end{equation}
The left-hand side of equation (\ref{EX-gf-3}) can be calculated from the definition of the fractional first derivative (\ref{app:A-1})
\begin{align}
\mathcal{F}\left\{\left(-\Delta_t^{(1)}\right)^{\frac{\alpha}{2}} x_a (t) ; \omega\right\}  = -i \omega|\omega|^{\alpha-2} \mathcal{F}\left\{x_a (t) ; \omega\right\}
= -i \omega|\omega|^{\alpha-2} \tilde{x}_a (\omega)
\, .
\label{EX-gf-4}
\end{align}
Thus, from equations (\ref{EX-gf-3}) and (\ref{EX-gf-4}), we obtain
\begin{equation}
-i m_\alpha \omega|\omega|^{\alpha-2} \tilde{x}_a (\omega) = \tilde{F}_a(\omega)
\, .
\label{EX-gf-5}
\end{equation}
Therefore, the coordinate functions in the conjugate variable $\omega$ are given by
\begin{equation}
\tilde{x}_a (\omega) = \frac{i \tilde{F}_a (\omega)}{m_\alpha \omega|\omega|^{\alpha-2}}
\, .
\label{EX-gf-6}
\end{equation}
Applying the inverse Fourier transformation, defined by the following convention,
\begin{equation}
f (t) = \frac{1}{\sqrt{2 \pi}} \int_{-\infty}^{+\infty} d \omega \tilde{f}(\omega) e^{-i \omega t}
\, ,
\label{EX-gf-7}
\end{equation}
we obtain
\begin{equation}
x_a (t) = \frac{i}{m_\alpha \sqrt{2 \pi}} \int_{-\infty}^{+\infty} \frac{d \omega \tilde{F}_a (\omega) e^{-i \omega t}}{\omega|\omega|^{\alpha-2}}
\, .
\label{EX-gf-8}
\end{equation}
Equation (\ref{EX-gf-8}) provides the general formula for calculating the coordinates of the fractional particle under the influence of a generalized force of the form given in equation (\ref{EX-gf-2}).

To exemplify its application, consider a fractional particle under the influence of a force of the following form
\begin{equation}
\mathcal{F}(t) = \hat{n} \sqrt{\frac{2}{\pi}} \sin \left(\frac{\pi \alpha}{2}\right) \Gamma(\alpha-1)|t|^{1-\alpha}
\, ,
\label{EX-gf-ex-9}
\end{equation}
where $\hat{n}$ is an $n$-dimensional vector, which we take to be unitary for convenience, i.e., $\lvert \hat{n} \rvert = 1$. Then, the coordinates are given by the following singular integral
\begin{equation}
x_a (t) = \frac{i \hat{n}_a}{m_\alpha \sqrt{2 \pi}} \int_{-\infty}^{+\infty} \frac{d \omega e^{-i \omega t}}{\omega}
\, .
\label{EX-gf-ex-10}
\end{equation}
The above integral can be solved by applying the residue theorem, yielding the result
\begin{equation}
x_a (t) = \frac{1}{2 m_{\alpha}} \hat{n}_{a}
\, .
\label{EX-gf-ex-11}
\end{equation}
From this equation, we can see that the force from equation (\ref{EX-gf-ex-9}) generates constant coordinate distributions. As discussed in the previous section, these distributions have null momenta and kinetic energy.

\section{Discussions}\label{sec:disc}

In this paper, we have proposed a model of a fractional particle with a first-order fractional derivative. This model generalizes the standard particle by introducing a kinetic energy term quadratic in the fractional derivative of coordinate distributions. The kinetic energy of the fractional particle with a first-order derivative is also quadratic in fractional momenta, contrasting with the fractional particle with a fractional Laplacian studied previously in \cite{Vancea:2023tpb,Vancea:2023cc}. We discussed simple systems with fractional particles possessing vanishing fractional linear momenta and kinetic energy, demonstrating that the standard free particle kinematics are reproduced in the classical limit. However, it is important to note that the classical model is local, and the value of the fractional parameter for which the classical limit is obtained lies outside the admissible range.

We provided the general solution for a particle subjected to generalized fractional non-local forces with fractional first derivatives and showed that a force of this type generates constant coordinate distributions. These types of systems admit analytic solutions. In the example of particles under generalized forces, several interesting physical and mathematical properties emerge. 

The generalized force $f_a(x, t)$ in equation (\ref{EX-gf-2}) introduces a non-local interaction in time. This non-locality is related to the fractional first derivative, and reflects a new property when compared with  classical forces which are typically local in time. This property suggests potential applications in modeling systems where memory effects or long-range temporal correlations are significant.

The transition to the frequency domain via the Fourier transform in equations (\ref{EX-gf-3}) to (\ref{EX-gf-6}) shows how the generalized forces impact the system and emphasize the frequency-dependent nature of the response, where different frequencies can be affected differently by the fractional dynamics. This information could be particularly useful in understanding the behavior of the system under periodic or quasi-periodic forces.

The result that the force in equation (\ref{EX-gf-ex-9}) leads to constant coordinate distributions equation (\ref{EX-gf-ex-11}) can be interpreted physically. It suggests that such generalized forces could model situations where the particle reaches a steady state or equilibrium position, despite the non-local nature of the interaction. This contrasts to classical mechanics, where a constant force typically results in linear acceleration. The constant coordinate distributions imply zero kinetic energy and momentum. This result is non-trivial and suggests that the system's energy is not stored in kinetic forms but possibly in the potential or interaction terms. This shift in energy distribution could help understanding energy transfer and storage in fractional systems.

Several important problems remain to be addressed. Firstly, understanding the mathematical structure of the fractional momenta and kinetic energy is essential. Compared to previous models from \cite{Vancea:2023tpb,Vancea:2023cc}, the fractional particle with a first-order derivative has the advantage of a quadratic structure, suggesting the existence of a bilinear form. However, the fractional derivatives used here are non-local and defined by singular integrals, lacking an obvious connection with locally defined objects. Since fractional first derivatives are less understood, studying the fractional particle model provides a strong motivation to further study these derivatives.

Secondly, it is essential to study more complex systems and models with non-local potential energies and more general force distributions. These studies could reveal other non-trivial properties and help us understand the dynamics of such models better. We hope to report on these topics in the future.

\section*{Acknowledgments}

C. F. L. G. and I. V. V. acknowledge J. Weberszpil  and M. C. Rodriguez for discussions. I. V. V. received partial support from the Basic Research Grant (APQ1) from the Carlos Chagas Filho Foundation for Research Support of the State of Rio de Janeiro (FAPERJ), grant number E-26/210.511/2024. 
 C. F. L. G. also acknoledges FAPERJ for partial support through an APQ1 grant.

\section*{Appendix A: Basic Properties of Fractional First Derivative}
\renewcommand{\theequation}{A.\arabic{equation}}
\setcounter{equation}{0} 

In this Appendix, we present some basic properties of the fractional first derivative that are used throughout this paper. More details can be found in \cite{Pozrikidis:2016}.

The fractional first derivative was defined in equations (\ref{FPFD-der-1}) and (\ref{FPFD-der-2}) repeated here for convenience
\begin{equation}
\left(-\Delta_t^{(1)}\right)^{\frac{\alpha}{2}} f(t)
= c_\alpha^{(1)} \text{P.V.} \int_{-\infty}^{+\infty} \frac{\tau f(t+\tau)}{|\tau|^{1+\alpha}} 
d \tau
\, ,
\label{app:A-1}
\end{equation}
where $\alpha \in [0, 2)$ and the constant is expressed in terms of Euler Gamma functions
\begin{equation}
c_\alpha^{(1)}=\frac{2^{\alpha-1}}{\sqrt{\pi}} \frac{\Gamma\left(\frac{1+\alpha}{2}\right)}{\Gamma\left(\frac{2-\alpha}{2}\right)}
\, .
\label{app:A-2}
\end{equation}
The functions $f$ belong to the space $\mathcal{X}$ on which the operator 
$\left(-\Delta_t^{(1)}\right)^{\frac{\alpha}{2}}$ is applied with proper behavior at $t=0$. The fractional first derivative is related to the one dimensional fractional Laplacian $\left(-\Delta_t \right)^{\frac{\alpha}{2}}$ by a standard first order derivative, hence its name, which can be formally written as
\begin{equation}
\left(-\Delta_t \right)^{\frac{\alpha}{2}} = \frac{d}{d t} \left(-\Delta_t^{(1)}\right)^{\frac{\alpha}{2}}
\, .
\label{app:A-3}
\end{equation}  
This shows that the fractional first derivative is also related to the Riesz derivative $\left(-\Delta_t \right)^{-\frac{\alpha}{2}}$ by
\begin{equation}
- \frac{d}{d t} \left(-\Delta_t^{(1)}\right)^{\frac{\alpha}{2}} \left(-\Delta_t \right)^{- \frac{\alpha}{2}} = \mathds{1}
\, .
\label{app:A-4}
\end{equation}
The above relations hold on spaces of functions $\mathcal{X}$ with sufficient regular properties. In general, in order to apply different relations of fractional calculus to fractional first derivative and fractional Laplacian together, care must be taken since equivalent formulations of these operators
are required. For fractional Laplacian, these equivalent forms have been shown to hold on $\mathcal{L}^p, p \in[1, \infty)$, $\mathcal{C}_0$ or $\mathcal{C}_{b u}$ according to the Theorem 1.1 from \cite{Kwasnicki:2017}, where
$\alpha \in (0, 2)$, $p \in [1, \frac{1}{\alpha})$, $\mathcal{L}^p$ denotes the Lebesgue spaces, $\mathcal{C}_0$ denotes the space of the continuous functions and  $\mathcal{C}_{b u}$ is the space of bounded uniformly continuous functions.  

An alternative formulation of the fractional first derivative can be given in terms of an integral over $\mathbb{R}^{\star}_{+}$ only. The corresponding formula is
\begin{equation}
(-\Delta^{(1)}_{t})^{\frac{\alpha}{2}} f(t)= c_\alpha^{(1)} \int_0^{+\infty} \frac{f(t+\tau)-f(t-\tau)}{\tau^\alpha} d \tau 
\, ,
\label{app:A-5}
\end{equation}
where the functions $f(t)$ must be well behaved on $\mathbb{R}^{\star}_{+}$ so that the integral make sense.

\section*{Appendix B: Calculation of Divergent Integral}
\renewcommand{\theequation}{B.\arabic{equation}}
\setcounter{equation}{0} 

Here, we calculate the divergent integral from formula (\ref{FPFD-sym-tran-3}). Given the principal value
\begin{equation}
I_{\alpha} = \text{P.V.} \int_{-\infty}^{+\infty} \frac{\tau \, d\tau}{|\tau|^{1+\alpha}}
\, ,
\label{app:B-1}
\end{equation}
we split it as
\begin{align}
I_{\alpha} & = \text{P.V.} \left( \int_{-\infty}^{0} \frac{\tau}{|\tau|^{1+\alpha}} \, d\tau + \int_{0}^{+\infty} \frac{\tau}{|\tau|^{1+\alpha}} \, d\tau \right)
\nonumber
\\
& = \text{P.V.} \left( \frac{1}{(-1)^{1+\alpha}} \int_{-\infty}^{0} \tau^{-\alpha} \, d\tau + \int_{0}^{+\infty} \tau^{-\alpha} \, d\tau \right)
\label{app:B-2}
\, .
\end{align}
By following the principal value prescription, we introduce symmetric cutoffs around 
$\tau = 0$ 
\begin{equation}
I_{\alpha} = \lim_{\epsilon \to 0} \left( \frac{1}{(-1)^{1+\alpha}} \int_{-\infty}^{-\epsilon} \tau^{-\alpha} \, d\tau + \int_{\epsilon}^{+\infty} \tau^{-\alpha} \, d\tau \right)
\, .
\label{app:B-3}
\end{equation}
We compute each integral from the right hand side of the equation (\ref{app:B-3}) separately.

For $\tau > 0$ and $\alpha \neq 1$, we have
\begin{equation}
\int_{\epsilon}^{+\infty} \tau^{-\alpha} \, d\tau = \left. \frac{\tau^{1-\alpha}}{1-\alpha} \right|_{\epsilon}^{+\infty} = \lim_{b \to \infty} \left( \frac{b^{1-\alpha}}{1-\alpha} - \frac{\epsilon^{1-\alpha}}{1-\alpha} \right).
\label{app:B-4}
\end{equation}

\textit{Case I}: For $0 < \alpha < 1$, the term $\frac{b^{1-\alpha}}{1-\alpha}$ diverges as $b \to \infty$, hence this integral (\ref{app:B-4}) diverges.

\textit{Case II}: For $1 < \alpha < 2$, the integral is 
\begin{equation}
\int_{\epsilon}^{+\infty} \tau^{-\alpha} \, d\tau = - \frac{\epsilon^{1-\alpha}}{1-\alpha}
\, .
\label{app:B-5}
\end{equation}

For $\tau < 0$ and $\alpha \neq 1$, we have
\begin{equation}
\int_{-\infty}^{-\epsilon} \tau^{-\alpha} \, d\tau = \left. \frac{\tau^{1-\alpha}}{1-\alpha} \right|_{-\infty}^{-\epsilon} = \lim_{a \to -\infty} \left( \frac{a^{1-\alpha}}{1-\alpha} - \frac{(-\epsilon)^{1-\alpha}}{1-\alpha} \right)
\, .
\label{app:B-6}
\end{equation}

\textit{Case I}: For $0 < \alpha < 1$, the term $\frac{a^{1-\alpha}}{1-\alpha}$ diverges as $a \to -\infty$, which makes this integral divergent.

\textit{Case II}: For $1 < \alpha < 2$, the integral is
\begin{equation}
\int_{-\infty}^{-\epsilon} \tau^{-\alpha} \, d\tau =  \frac{(-\epsilon)^{1-\alpha}}{1-\alpha}
\, .
\label{app:B-7}
\end{equation}

By using the results from the equations (\ref{app:B-5}) and (\ref{app:B-7}) into the equation (\ref{app:B-3}), the principal value integral can be computed as
\begin{equation}
I_{\alpha} = \lim_{\epsilon \to 0} 
\left[ 
	\frac{1}{(-1)^{1+\alpha}} \left( \frac{(-\epsilon)^{1-\alpha}}{1-\alpha} 	\right) - \frac{\epsilon^{1-\alpha}}{1-\alpha} 
\right]
= 
\frac{1}{1-\alpha}\lim_{\epsilon \to 0} 
\left[ 
	(-1)^{-2 \alpha} \epsilon^{1-\alpha} - \epsilon^{1-\alpha} 
\right] = 0
\, .
\label{app:B-8}
\end{equation}


\end{document}